\documentclass[11pt]{article} 
\usepackage{graphicx,amssymb,epsf,rotate} 
\begin{document}
\begin{center}
{\Large \bf
Kinetic mixing and symmetry breaking dependent interactions 
of the dark photon
}
\vskip 1cm
\renewcommand{\thefootnote}{\fnsymbol{footnote}}
Biswajoy Brahmachari$^1$\footnote{biswa.brahmac@gmail.com} and 
Amitava Raychaudhuri$^2$\footnote{palitprof@gmail.com}\\
\vskip 1cm
(1) Department of Physics, 
Vidyasagar Evening College, \\
39 Sankar Ghosh Lane, Kolkata 700006, India\\
\vskip .5cm
(2) Department of Physics, 
University of Calcutta, \\ 
92 Acharya Prafulla Chandra Road, Kolkata 700009, India
\end{center}
\vskip .5cm
\begin{center}
\underbar{\bf Abstract}
\end{center}

We examine spontaneous symmetry breaking of  a renormalisable
$U(1) \times U(1)$ gauge theory coupled to fermions when kinetic
mixing is present.  We do not assume that the kinetic mixing
parameter is small. A rotation plus scaling is used to remove the
mixing  and put the gauge kinetic terms in the canonical form.
Fermion currents are also rotated in a non-orthogonal way by this
basis transformation.  Through suitable redefinitions the interaction
is cast into a diagonal form.  This framework, where mixing
is absent, is used for subsequent analysis.  The
symmetry breaking determines the fermionic current which couples
to the massless gauge boson. The strength of this coupling as
well as the couplings of the massive gauge boson are extracted.
This  formulation is used to  consider a  gauged
model for dark matter by identifying the massless gauge boson
with the  photon and the massive state to its dark counterpart.
Matching the coupling of the residual symmetry with that of
the photon sets a {\em lower} bound on the kinetic mixing
parameter. We
present analytical formulae of the couplings of the dark photon
in this model and indicate some physics consequences.
\parindent 0pt


\texttt{Key Words:~~Kinetic mixing, Dark Matter } 

\renewcommand{\thesection}{\Roman{section}} 
\setcounter{footnote}{0} 
\renewcommand{\thefootnote}{\arabic{footnote}} 
\noindent

\section{Introduction}

In a non-abelian gauge theory the field tensor $F_{\mu\nu}^a$ is
gauge covariant and the kinetic term is ${\cal L} =
-\frac{1}{4} F^{a \mu\nu} F_{\mu\nu}^a$, where $\mu, \nu$ are
Lorentz indices and $a$ is a gauge index both of which are summed
over. This form is determined by Lorentz and gauge invariance.
For a $U(1)$ gauge theory, on the other hand, $F^{\mu \nu}$ by
itself is gauge invariant. Therefore, if there are several
$U^i(1) ~(i = 1, \ldots n)$ factors in a theory the possibility
of mixed terms in the Lagrangian of the form $-\alpha_{ij} F^{i
\mu\nu} F_{\mu\nu}^ j ~(i \neq j)$, where $\alpha_{ij}$ quantify
the mixing (in a given basis),  opens up.  Indeed, such {\em
kinetic mixing} has been noted in the literature  \cite{kinetic}
and its origin, especially in the context of grand unification
where two $U(1)$ factors are often encountered, examined
\cite{gut}. If both the $U(1)$ are embedded in a grand unified
theory (GUT) such as $E_6$ then at the unification scale the
mixing will vanish but it could be generated at low energy where
the GUT symmetry is not exact, by renormalisation group (RG) effects.
Phenomenological applications in the context of dark matter have
considered the kinetic mixing of a ``dark sector" $U(1)$ gauge
field with the $U(1)_Y$ of the standard model \cite{Ymixth}. The
detectability of a such a dark photon in a number of experiments
using different approaches has
been examined \cite{Ymixex}.  The alternative of the kinetic
mixing of the  ``dark sector" $U(1)$  with $U(1)_{EM}$ has
also been proposed \cite{pmixth}. Possible tests of such a
photon-dark photon mixing scenario are available in the
literature \cite{pmixex}. Consequences of mixing between several dark
sector $U(1)$ factors have been illustrated in \cite{genmix}.

In this work our endeavour has been to take a detailed look at
the effect of spontaneous symmetry breaking on a theory with two
kinetically mixed $U(1)$ factors where the gauge bosons also
couple to fermions, that is, when interaction terms are present.
In the literature such models are  usually analysed with the
assumption that the coefficient of the mixing term, $c$, is
small. We have not imposed this restriction\footnote{For a
discussion of kinetic mixing of a dark photon with the
hypercharge $U(1)_Y$ without the small mixing restriction see, for
example, \cite{GGRoss}.}.  On the contrary, we show that
depending on the two $U(1)$ gauge coupling strengths, $g_1$ and
$g_2$,  a {\em lower} bound on the magnitude of $c$ will exist if
the coupling of the final unbroken gauge symmetry is to match
that of electromagnetism, $e$.  In particular, we show that  in
this case $c$ must satisfy
\begin{equation}
\frac{1}{4}\left|\frac{g^2_1 + g^2_2}{2 e^2} -  1 \right| \leq
|c| \leq \frac{1}{4} \;.
\label{bound}
\end{equation}
In the special case $g_1=g_2=e$ the lower bound on $c$ becomes
zero. Also, if $(g_1^2 + g_2^2) >  4 e^2$ then there is no solution.

In general the presence of two  $U(1)$ symmetries will entail all
particles to carry two distinct charges. Consequently there are
two fermionic currents. These currents couple exclusively
to the two gauge bosons of the theory without  any 
cross terms. In this way the starting basis of our analysis is defined.
 
We proceed  in the following stages. In
the first the mixing term $F^1_{\mu\nu} F^{2\mu \nu}$ is removed
by a transformation of gauge bosons involving an orthogonal
rotation and a scaling\footnote{The scaling of the gauge
fields is matched by an inverse scaling of the corresponding
couplings.}.  The initial basis where kinetic mixing terms  are
present is denoted as the $A$ basis and the second where
non-diagonal terms are removed as the $B$ basis.  Thereafter
spontaneous symmetry breaking of the type, $U^1(1)\times U^2(1)
\rightarrow U^3(1)$ takes place. This causes a further orthogonal
rotation of gauge bosons taking the $B$ basis to the mass
eigenstates which we term as the $X$ basis. One of the states,
$X^1_\mu$, associated with  the unbroken $U^3(1)$ remains
massless while the other state $X^2_\mu$ is massive. 
These mass eigenstates form an
orthonormal basis.  The original $A$ basis where mixing terms are
present and which is related to this mass basis $X$ by
orthogonal and scaling transformations cannot then be
orthogonal.  That leads us to no conflict as we can  define
charges of fermions and scalars consistently in the $B$ basis
which  {\em is} orthogonal.

We evaluate the couplings of the massless and massive gauge boson
states to fermions after symmetry breaking. The massless
eigenstate, $X^1_\mu$, couples to one particular combination of
the two fermion currents.  We express this specific combination
in terms of the direction of symmetry breaking in the $U^1(1)
\times U^2(1)$ space and  further determine the coupling strength
of the massless boson to the current  related to the
unbroken charge.  The coupling of the
massive gauge boson, $X^2_\mu$, is conveniently given in terms
of the above current combination and another 
current which is orthogonal to it.  We observe that the 
coupling of  $X^2_\mu$ to the unbroken combination
is controlled by the kinetic mixing parameter $c$. 

We have indicated how these results on couplings may offer a
window on the physics of Dark Matter via a {\em calculable}
ordinary matter-dark matter interaction  strength. We have given
two examples. In the first, ordinary matter does not have any
dark charges and also, as expected, the dark matter does not have
electric charge. In the second example, an extra $U(1)$ of the
dark sector is kinetically mixed with normal $U(1)_{EM}$.
Spontaneous symmetry breaking occurs in such a manner that  the
unbroken direction remains along $U(1)_{EM}$. This means that
only the dark $U(1)$ is broken. In such an event due to the
presence of kinetic mixing in the unbroken theory we derive
relations between gauge couplings in the broken theory.  Such
relations will not exist if kinetic mixing is absent.

Our paper is arranged as follows. In the next section we set 
up the notation and the transformation
from the $A$- (mixed) to the $B$- (unmixed) basis. Symmetry
breaking is considered in the following section and the $X$-basis
is defined  as the (orthogonal) mass basis for gauge bosons. Analytic 
expressions for the couplings of the gauge
boson mass eigenstates to fermions are given in the next section.
Possible application of these ideas in the context of dark matter
are then considered. We end with our conclusions.

\section{Removing kinetic mixing by rotation and scaling}

In general, the kinetic terms for a gauge theory consisting of
two  $U(1)$ groups can be written as\footnote{The theory may have
other non-abelian gauge symmetries which we suppress.}
\begin{equation}
{\cal L}_{gauge}= -\frac{1}{4} ~F^1_{\mu \nu} F^{1~\mu \nu} 
- \frac{1}{4} ~F^2_{\mu \nu} F^{2~\mu \nu} - 2c~F^1_{\mu \nu}
F^{2~\mu \nu} \;. 
\label{gkm0}
\end{equation}
Here the field strengths are expressed in terms of gauge fields by
the usual formula
\begin{equation}
F^r_{\mu \nu}= \partial_\mu A^r_\nu - \partial_\nu
A^r_\mu,~~~~{\rm where}~~r=1,2,
\end{equation}
and $c$ is a real kinetic mixing parameter. Gauge invariance  cannot fix
the magnitude of  $c$.  In fact $c$ has different values for different
basis choices. Once we are able to fix the 
basis $A^1_\mu,A^2_\mu$ using a set of
physical arguments, then $c$ becomes a meaningful parameter.
The Lagrangian for the interaction of fermions with gauge bosons is
\begin{equation}
{\cal L}_{int} = g_1 j^\mu_1 A^1_{\mu} + g_2 j^\mu_2 A^2_{\mu} =
\pmatrix{j^\mu_1 & j^\mu_2} \pmatrix{g_1 & 0 \cr 0 & g_2}
\pmatrix{A_\mu^1 \cr A_\mu^2}.   
\label{int1}
\end{equation}
Above, $j^\mu_r ~~(r=1,2)$, is the fermionic current due to the
presence of $U(1)_r$ charge
\begin{equation}   
j_r^\mu = q_r^f \bar{\psi}  \gamma^\mu \psi \;\; .
\label{fcurr0}
\end{equation}             
$g_i$ is the corresponding coupling strength.  We see
that the initial basis, called $A$ basis, is fixed by demanding that
couplings of gauge bosons to fermions is diagonal.

Through an orthogonal rotation by $\pi/4$  in the
$A_\mu^1-A_\mu^2$ sector\footnote{This rotation angle is
determined by the equality of the coefficients of the two
diagonal terms, $F^i_{\mu \nu} F^{i~\mu \nu} ~(i =1,2)$,  in Eq.
(\ref{gkm0}) and is independent of  the magnitude of $c$. In
the absence of $c$ the rotation through {\em any} angle would
yield an equivalent basis.} followed by  scaling, by a  factor
which can always be chosen to be real,   one can remove the
kinetic mixing and bring the gauge Lagrangian in Eq.
(\ref{gkm0}) to the canonical form.  After these transformations
one has
\begin{equation}
{\cal L}_{gauge} =  -{1 \over 4}~ {G}^1_{\mu \nu} {G}^{1~\mu \nu} 
- {1 \over 4}~ {G}^2_{\mu \nu} {G}^{2~\mu \nu} \;\;.
\label{lag1}
\end{equation}
Here the redefined field strength tensors are
\begin{equation}
G^r_{\mu \nu}= \partial_\mu {B}^r_\nu - \partial_\nu
{B}^r_\mu,~~~~{\rm where}~~r=1,2.
\end{equation}
The new basis is defined by the transformation equation
\begin{equation}
 \pmatrix{A^1_\mu \cr ~A^2_\mu} = 
\frac{1}{2\sqrt{2}} \pmatrix{\sqrt{1 \over \lambda_{1}}  & 
-\sqrt{1 \over \lambda_{2}}  \cr   
\sqrt{1 \over \lambda_{1}} &
\sqrt{1 \over \lambda_{2}}  }\pmatrix{{B}^1_\mu \cr {B}^2_\mu}.
\label{rs1}
\end{equation}
In this new basis, here termed the $B$ basis, there is no kinetic 
mixing, but we have lost the diagonal form of interaction with fermions.
In  the transformation matrix given in Eq. (\ref{rs1}), the
parameters $\lambda_1,\lambda_2$ are given by 
\begin{equation}
\lambda_{1,2} = {\frac{1}{4} \pm   c}.  
\label{lam}
\end{equation}
We observe that $\lambda_1,\lambda_2$ are the eigenvalues of a real
symmetric matrix formed by the coefficients of terms  
in Eq. (\ref{gkm0}).  If the scaling transformations 
in Eq. (\ref{rs1}) are real then $\lambda_1$ and $\lambda_2$ must 
be positive, which results in  the following
inequalities\footnote{We show later that physical processes are
well defined in the $c \rightarrow 1/4$ limit.}:
\begin{equation}
|c| < \frac{1}{4} \;\;,\;\; 0 < \lambda_{1,2} < \frac{1}{2} \;\;,\;\;
{\rm subject ~to}\;\;\lambda_1 + \lambda_2 = \frac{1}{2} \;\;.
\label{bounds}
\end{equation}
Under the transformation $c \leftrightarrow -c$ we get
$\lambda_1 \leftrightarrow \lambda_2$. So, we
can keep $c > 0$ and $\lambda_1 > \lambda_2$ in this analysis
with the understanding that the results for negative $c$ can be
obtained by the prescription noted above.

It is to be borne in mind  that Eq. (\ref{rs1}) is not an
orthogonal transformation so if the $B$ basis is
orthogonal\footnote{It is shown in the following  that the $B$ basis is
related to the physical mass eigenstates by an orthogonal
transformation.} the $A$ basis is not.  
We will define the $U(1) \times U(1)$ charges in the
$B$ basis which is orthonormal  keeping in
mind that in this basis off-diagonal interactions with fermions are
present.  However, the mixing parameter $c$, defined
in the $A$ basis can still be constrained as we discuss 
now.

Let us  rewrite Eq. (\ref{int1}) as
\begin{equation}           
{\cal L}_{int} = 
\frac{1}{2\sqrt{2}} \pmatrix{j^\mu_1 & j^\mu_2}
\pmatrix{g_1 \over \sqrt{\lambda_{1}}  & -g_1 \over \sqrt{\lambda_{2}}  \cr 
g_2 \over \sqrt{\lambda_{1}} & g_2 \over \sqrt{\lambda_{2}}  }
\pmatrix{{B}^1_\mu \cr {B}^2_\mu}.
\label{int2n}
\end{equation}                              
It is amply evident now that these are two equivalent formulations 
of the same phenomenon. The first description corresponds to 
Eq. (\ref{gkm0}) and Eq. (\ref{int1}) where there is kinetic mixing 
among the $U(1)$ gauge field strengths (i.e., $c \neq 0$) and the currents 
couple only to the corresponding gauge bosons, i.e., $j_r^\mu$ to $A_\mu^r$
for $r = 1,2$. In the second picture given by Eq. (\ref{lag1})
and Eq. (\ref{int2n}) there is no kinetic mixing among gauge boson
fields, ${B}_\mu^{1,2}$, but the currents ${j}_{1,2}^\mu$ couple 
to both gauge bosons. It is to be noted that in Eq.
(\ref{int2n}) a change of
$c$ only affects the scaling within the matrix and in the limit $c
\rightarrow 0$ we have
\begin{equation}           
{\cal L}_{int} = 
\frac{1}{\sqrt{2}} \pmatrix{j^\mu_1 & j^\mu_2}
\pmatrix{g_1  & -g_1   \cr 
g_2  & g_2   }
\pmatrix{{B}^1_\mu \cr {B}^2_\mu}.
\label{int2b}
\end{equation}                              
From Eq. (\ref{rs1}) we can easily see that in this limit
$B^1_\mu$ and $B^2_\mu$ have equal admixtures of $A^1_\mu$ and
$A^2_\mu$.  This result is reminiscent of degenerate perturbation
theory.

An alternate but useful way of rewriting interactions in
Eq.  (\ref{int2n})  is to express it in terms of a redefined
set of  currents $J_{1,2}^\mu$ which couple  diagonally to
the gauge bosons ${B}_\mu^{1,2}$. One then has
\begin{equation}           
{\cal L}_{int}  
= \pmatrix{{J}^\mu_1 & {J}^\mu_2} 
\pmatrix{\tilde{g}_1 & 0 \cr 0 & \tilde{g}_2}
\pmatrix{{B}_\mu^1 \cr {B}_\mu^2} \;.
\label{int2}
\end{equation}                              
In this process, currents involving fermions are scaled 
and rotated now, as, 
\begin{equation}
\pmatrix{J^\mu_1 \cr J^\mu_2} 
= \pmatrix{\cos\phi  & \sin\phi \cr  -\cos\phi  & \sin\phi}
\pmatrix{ j^\mu_1 \cr j^\mu_2},
\label{curr1}
\end{equation}
which is a non-orthogonal transformation, and
\begin{equation}
\cos \phi =  \frac{g_1}{\sqrt{g_1^2 + g_2^2}},
\;\; \sin \phi =  \frac{g_2}{\sqrt{g_1^2 + g_2^2}},
\;\;\tilde{g}_{1} =  \frac{\sqrt{g_1^2 +
g_2^2}}{2\sqrt{2\lambda_{1}}},
\;\; \tilde{g}_{2} =  \frac{\sqrt{g_1^2 + g_2^2}}{2\sqrt{2\lambda_{2}}}\;.
\label{coup1}
\end{equation} 
For $c > 0$ we have $\lambda_1 > \lambda_2$ which leads to 
$\tilde{g}_{2} > \tilde{g}_{1}$.

In the special case $g_1 = g_2 \equiv g$, which we consider in an
example later, the relations in Eq. (\ref{coup1}) become 
\begin{equation}
\cos \phi =   \sin \phi =  \frac{1}{\sqrt{2}},
\;\;\tilde{g}_{1} =  \frac{g}{2\sqrt{\lambda_{1}}},
\;\; \tilde{g}_{2} =  \frac{g}{2\sqrt{\lambda_{2}}}\;.
\label{coup1degen}
\end{equation} 

It is to be noted that though Eq.  (\ref{int2}) bears a strong
resemblance to Eq.  (\ref{int1}) a major difference is that the
currents $j^\mu$ and $J^\mu$ are related by a non-orthogonal
rotation.

\section{Spontaneous symmetry breaking}
At this point we are in a position to
consider symmetry breaking of  the $U^1(1) \times U^2(1)$
theory.  For  a scalar field $\Phi$ with $U^{1,2}(1)$ charges $q_{1,2}^s$ 
the covariant derivative is
\begin{eqnarray}                                                   
D^\mu \Phi &=& \left[\partial^\mu - i g_1 q_1^s A_1^\mu - i g_2 q_2^s
A_2^\mu \right] \Phi \nonumber \\
 &=& \left[ \partial^\mu - i \tilde{g}_{1}  {Q}_1^s B_1^\mu  
- i  \tilde{g}_{2}  {Q}_2^s B_2^\mu \right] \Phi     \;\;.
\label{cov1}
\end{eqnarray}
In our convention, we have assigned charges $Q_i$ in the $B$ basis
and the $q_i$ charges are in the $A$ basis. They are related
through Eq.   (\ref{curr1}). Thus
\begin{equation}
\pmatrix{Q_1 \cr Q_2}=         
         \pmatrix{\cos\phi & \sin\phi \cr   
         -\cos\phi  & \sin\phi }
\pmatrix{q_1 \cr q_2}\;.
\label{coup1a}
\end{equation} 
We can now consider spontaneous breaking of the $U^1(1) \times U^2(1)$
symmetry by the scalar field  developing a vacuum expectation
value  $ \langle \Phi \rangle =
v/\sqrt{2} \ne 0$. The gauge boson mass matrix in the $B_{1,2}$ basis 
is
\begin{equation}
M^2_{gauge}= v^2 \pmatrix{(\tilde{g}_{1} {Q}_1^s)^2 & 
(\tilde{g}_{2} {Q}_2^s) (\tilde{g}_{1} {Q}_1^s) \cr 
(\tilde{g}_{2} {Q}_2^s) (\tilde{g}_{1} {Q}_1^s) &
(\tilde{g}_{2} {Q}_2^s)^2 }.
\end{equation}
The mass eigenstates are denoted by  $X^1_\mu,X^2_\mu$. One
eigenstate has a zero eigenvalue while the other one is
massive\footnote{The complex scalar field $\Phi$ provides the
longitudinal mode for $X^2_\mu$ and also results in a real scalar
boson. The latter can couple to the SM Higgs boson
through quartic terms in the scalar potential leading to a `Higgs
portal' for the dark matter \cite{Wilczek}.}.
Because $X^1_\mu$ and $X^2_\mu$ are eigenvectors of a real
symmetric matrix with distinct eigenvalues, they are orthogonal.
Furthermore, we know that the diagonalizing matrix is an
orthogonal matrix.
\begin{equation}
\pmatrix{X^1_\mu \cr X^2_\mu} = \pmatrix{\cos \theta  & -\sin \theta 
\cr \sin \theta & \cos \theta} \pmatrix{ B^1_\mu \cr  B^2_\mu}.  
\label{orthr}
\end{equation}
The mixing angle is
\begin{equation}
\cos\theta = \frac{1}{N} ~\left[\tilde{g}_{2} {Q}_2^s \right], \;\;
\sin\theta = \frac{1}{N} ~\left[\tilde{g}_{1} {Q}_1^s \right],
\label{angles}
\end{equation}
where the normalization factor is given by, 
\begin{equation}
N^2 = \left(\tilde{g}_{2} {Q}_2^s \right)^2
+ \left(\tilde{g}_{1} {Q}_1^s \right)^2.
\label{norm}
\end{equation}
The two eigenvalues of the mass matrix are, 
\begin{equation}
m_1^2 = 0, ~~m_2^2 = N^2 v^2.
\label{masses}
\end{equation}
Interactions of the mass eigenstates $X^1_\mu$ and $X^2_\mu$ can now
be written neatly. From Eq. (\ref{int2}) one has 
\begin{equation}           
{\cal L}_{int} = 
\pmatrix{{J}^\mu_1 & {J}^\mu_2} 
\pmatrix{\tilde{g}_1 & 0 \cr 0 & \tilde{g}_2}
\pmatrix{\cos \theta  & \sin \theta 
\cr -\sin \theta & \cos \theta}   
\pmatrix{X^1_\mu \cr X^2_\mu} .
\label{int2a}
\end{equation}
When $U(1)\times U(1)$ symmetry is spontaneously broken to 
 a residual $U(1)$, there is an associated conserved charge 
which is a linear combination of $Q_1$ and $Q_2$. This
conserved charge can be written in a normalised form as
\begin{equation}
Q=\alpha_1 Q_1 + \alpha_2 Q_2, \;\;\; (\alpha_1^2 + \alpha_2^2 = 1)\;,
\label{chargeQ}
\end{equation}
such that the scalar  field $\Phi$ which acquires a vacuum
expectation value triggering the symmetry breaking 
satisfies\footnote{This direction is independent of whether we
choose the $A_{1,2}^\mu$ basis or the $B_{1,2}^\mu$ basis. Here
we have used the $B_{1,2}^\mu$ basis. }
\begin{equation}
\alpha_1 Q^s_1 + \alpha_2 Q^s_2=0. \label{sym0}
\end{equation}
This implies (up to an overall sign)
\begin{equation}
\alpha_1 = \frac{Q_2^s}{\sqrt{Q_1^{s2} + Q_2^{s2}}} \;\; , \;\; 
\alpha_2 = -\frac{Q_1^s}{\sqrt{Q_1^{s2} + Q_2^{s2}}} \;.
\label{angles2}
\end{equation}
One can also define another charge, which is not conserved, and is orthogonal 
in direction to $Q$:
\begin{equation}
Q^\prime=-\alpha_2 Q_1 + \alpha_1 Q_2.
\label{chargeQp}
\end{equation}
The non-conservation of $Q^\prime$ is due to the fact that the
corresponding $U(1)$ symmetry is broken.

One can use Eq. (\ref{angles2}) to express the mass of $X^2_\mu$
in terms of $\alpha_{1,2}$. From Eqs. (\ref{coup1}), (\ref{norm}), and
(\ref{masses}) one has 
\begin{equation}
m_2^2 = \frac{(g_1^2 + g_2^2)(Q_1^{s2} + Q_2^{s2})}{8}
\left(\frac{\alpha_1^2}{\lambda_2} + \frac{\alpha_2^2}{\lambda_1}\right) v^2.
\label{mass2}
\end{equation}

\section{Fermion interactions}
Recall that we had originally defined fermionic currents 
in Eq. (\ref{fcurr0}) in the presence of kinetic mixing. These currents 
had a diagonal interaction 
with gauge bosons in the $A$ basis. A combination
of $j^\mu_{1,2}$ was identified in Eq. (\ref{curr1}) to form
$J^\mu_{1,2}$ which had a diagonal
form of interaction in the $B$ basis. These currents will now
be further mixed during spontaneous symmetry breaking. We can express 
fermionic interactions of the gauge boson mass eigenstates in terms
of the currents defined through the charges $Q^f$ and $Q'^f$ as
\begin{equation}   
\hat{J}_1^\mu = Q^f \bar{\psi}  \gamma^\mu \psi = \left({J}^\mu_1
\alpha_1 + {J}^\mu_2 \alpha_2 \right) \;\;, \;\;
\hat{J}_2^\mu = Q'^f \bar{\psi}  \gamma^\mu \psi =
\left(-{J}^\mu_1 \alpha_2 + {J}^\mu_2 \alpha_1 \right) \;\; .
\label{fcurr1}
\end{equation}             
It is convenient to write the interaction Lagrangian of
massive and massless gauge bosons as
\begin{equation}   
{\cal L}_{int} = 
\sum_{i,j = 1,2} g_{ij} ~\hat{J}_i^\mu  ~X^j_\mu  \;.
\label{intdef}
\end{equation}
Here $X^1_\mu$ corresponds to the surviving $U(1)$  and
it couples only to $\hat{J}^\mu_1$. On the contrary $X^2_\mu$
couples to both $\hat{J}^\mu_1$ as well as the orthogonal
combination, namely, $\hat{J}^\mu_2$. To determine the coupling 
strengths $g_{ij}$ we reexpress Eq. (\ref{int2a}) as
\begin{equation}           
{\cal L}_{int} = 
\left[ \left\{\tilde{g}_1
{J}^\mu_1 \cos \theta - \tilde{g}_2 {J}^\mu_2 \sin \theta
\right\} X^1_\mu +
 \left\{\tilde{g}_1
{J}^\mu_1 \sin \theta + \tilde{g}_2 {J}^\mu_2 \cos \theta
\right\} X^2_\mu \right] \;.
\label{intx1x2}
\end{equation}
In particular, using Eqs. (\ref{angles}) and  (\ref{angles2}) the interaction 
of the massless gauge boson $X^1_\mu$ is
\begin{equation}           
{\cal L}_{X^1} = 
\frac{\tilde{g}_1 \tilde{g}_2}
{\sqrt{\tilde{g}_1^2 \alpha_2^{2} + \tilde{g}_2^2 \alpha_1^{2}}}
\left({J}^\mu_1 \alpha_1 + {J}^\mu_2 \alpha_2 \right)  X^1_\mu 
= 
~{g}_{11} ~\hat{J}^\mu_1 ~X^1_\mu \;.
\label{intx1}
\end{equation}
We can now read off the coupling strengths $g_{11}$ and $g_{21}$
from Eq. (\ref{intx1}).  We see that
\begin{equation}
{g}_{11} = \frac{\tilde{g}_1 \tilde{g}_2}
{\sqrt{\tilde{g}_1^2 \alpha_2^{2} + \tilde{g}_2^2 \alpha_1^{2}}},~~~ g_{21}=0 \;\;.
\label{def2}
\end{equation}
By a rearrangement of terms, one obtains the more 
familiar expression
\begin{equation}
\frac{1}{{g}_{11}^2} = 
\frac{\alpha_1^2}{\tilde{g}_1^2} + \frac{\alpha_2^2}{\tilde{g}_2^2}. \label{fam1}
\end{equation}
An interesting consequence of Eq. (\ref{fam1}) is that
\begin{equation}
\tilde{g}_1 \leq g_{11} \leq \tilde{g}_2.
\end{equation}
As $X^1_\mu$ corresponds to a surviving $U(1)$ symmetry, it couples
only to $\hat{J}^1_\mu$.  
The interaction of $X^2_\mu$ can be expressed as 
\begin{equation}           
{\cal L}_{X^2} = 
\frac{1}{\sqrt{\tilde{g}_1^2 \alpha_2^{2} + \tilde{g}_2^2 \alpha_1^{2}}}
\left[ - \alpha_1 \alpha_2
(\tilde{g}_1^2 - \tilde{g}_2^2)\hat{J}^\mu_1 
+ \left(\alpha_2^{2}\tilde{g}_1^2 + \alpha_1^{2}\tilde{g}_2^2 \right)
 \hat{J}^\mu_2 \right]  X^2_\mu, 
\label{intx2}
\end{equation}
whence, we can again read off the couplings of the heavy gauge boson $X^2_\mu$
with the two currents, viz.
\begin{eqnarray}           
g_{22} &=&  \sqrt{\tilde{g}_1^2 \alpha_2^{2} + \tilde{g}_2^2 \alpha_1^{2}}
=  \frac{\sqrt{g_1^2 +
g_2^2}}{2\sqrt{2}}\sqrt{\left(\frac{\alpha_2^2}{\lambda_1} + 
\frac{\alpha_1^2}{\lambda_2} \right)}\;\;, \label{x2coup22} \\
g_{12} &=&  - \frac{\alpha_1 \alpha_2 (\tilde{g}_1^2 - \tilde{g}_2^2)}
{\sqrt{\tilde{g}_1^2 \alpha_2^{2} + \tilde{g}_2^2 \alpha_1^{2}}}
= - \alpha_1 \alpha_2 \frac{\sqrt{g_1^2 +
g_2^2}}{2\sqrt{2}} \left(\frac{\lambda_2 -
\lambda_1}{\sqrt{(\lambda_1 \lambda_2)(\alpha_1^2
\lambda_1 + \alpha_2^2 \lambda_2)}}\right)\;\;.
\label{x2coup}
\end{eqnarray}

Here we  emphasise that $X^2_\mu$ couples to both  $\hat{J}^\mu_1$ 
as well as $\hat{J}^\mu_2$ because there is no symmetry which can force
it to couple to $\hat{J}^\mu_2$ only.

\section{Applications}

Even though the existence of dark matter was known for
a long time \cite{darkold}, in fact since the 1930's, recent satellite-based
experiments such as COBE and WMAP have brought
the issue to the foreground \cite{satel}. Analysis of
temperature anisotropies of Cosmic Microwave Background (CMB)
data found in the PLANCK experiment has shown that in the 
universe $26.8 \%$ of all matter and energy is dark matter 
\cite{satel2}. Dark Matter interacts with the visible
sector  by gravitational interactions. Other possible 
interactions of the dark sector with the visible one has to be tightly 
controlled in order for it to qualify as dark matter. Any such 
interaction, if it exists at all, cannot be stronger 
than the weak interactions, i.e., the candidate dark matter could
be a weakly interacting massive particle (WIMP) \cite{wimp}.
 Here we examine the possiblity of gauge kinetic mixing
between the ordinary photon and a dark counterpart being
at the origin of such an interaction. This  can also account for
the fact that halo properties of galaxies, studied in
cosmological simulations, hint towards dark matter
self-interactions \cite{sps,dama}.

The idea of a ``dark photon" kinetically mixed with the ordinary
photon has been  invoked before in theories of 
dark matter \cite{pmixth,pmixex}. In
these theories the dark matter (DM) coupling to the dark photon is of
comparable strength as the coupling of the ordinary photon to standard
model (SM) matter. Though the dark matter does not couple to the
ordinary photon, the SM matter develops a tiny coupling to the
dark photon through the small kinetic mixing. This leads to an
effective interaction between the dark and SM sectors whose
strength is controlled by kinetic mixing. In the subsections
below we examine how our calculations can be useful for such
considerations without the assumption of a small kinetic
mixing. In other words in the following discussions the
value of the mixing parameter is not necessarily small.

\subsection{Couplings and charges of dark photons}

In a realistic theory the  residual unbroken $U(1)$ group
has to be identified with $U(1)_{EM}$, i.e., the massless gauge
boson,  $X^1_\mu$, has to be the photon.  The immediate
consequence  of this identification is that $g_{11}$ is related
with the fine structure constant by 
\begin{equation}
g_{11} = e = \sqrt{4 \pi \alpha_{EM}} \;\;.
\label{alpha}
\end{equation}
As $g_{11}$ is now expressed in terms of $e$, we can rewrite Eq. (\ref{fam1})
as,
\begin{equation}
\alpha^2_1 \left({e^2 \over \tilde{g}^2_1}\right)
+ 
\alpha^2_2 \left({e^2 \over \tilde{g}^2_2}\right)=1. \label{fm2}
\end{equation}
This is a key equation, arising from the
identification of $X^1_\mu$ with the photon, which  
relates symmetry breaking parameters ($\alpha_{1,2}$) with the
kinetic mixing strength $c$ and gauge couplings $\tilde{g}_{1,2}$
of $X^{1,2}_\mu$ which are mass basis states.  Using Eq.
(\ref{coup1}) along with (\ref{lam}) one can rewrite it as:
\begin{equation}
\left(g_1^2 + g_2^2 \right) = 2 e^2 \left[1 + 4 c ~(\alpha_1^2 -
\alpha_2^2) \right] \;\;.
\label{fm2a}
\end{equation}

Eq. (\ref{fm2}) results in a {\em lower} bound on the kinetic
mixing parameter $c$. To see this, using Eq.  (\ref{coup1}) we
express the couplings
$\tilde{g}_1$ and $\tilde{g}_2$ in  units of $e$ as
\begin{equation}
\tilde{g}_i/e =  \frac{\sqrt{g_1^2 +
g_2^2}}{2\sqrt{2}e} \frac{1}{\sqrt{\lambda_i}} =
\sqrt{\frac{\xi}{\lambda_i}} \;\;\; (i = 1,2) \;\;. 
\label{scaled}
\end{equation}
Here we have defined a new quantity $\xi$:
\begin{equation}
\xi = \frac{1}{8}\frac{(g_1^2 + g_2^2)}{e^2} \;\;. 
\label{xi}  
\end{equation}
In terms of $\xi$, Eq. (\ref{fm2}) takes the shape of
\begin{equation}
\lambda_1 \alpha_1^2 + \lambda_2 \alpha_2^2 =  \xi \;\;.
\label{alphacr}  
\end{equation}
We may recall that $\lambda_{1,2}$ are determined by the kinetic
mixing parameter $c$ in Eq. (\ref{lam}). Using 
$\alpha_1^2+ \alpha_2^2 = 1$ one can solve for  $\alpha_{1,2}$,
\begin{equation}
\alpha_1  = \sqrt{\frac{c + (\xi -\frac{1}{4})}{2c}}   \;\;,\;\;
\alpha_2  = \sqrt{\frac{c - (\xi -\frac{1}{4})}{2c}}   \;.
\label{alphacr2}  
\end{equation}
Now since  $0 \le \alpha^2_1 \le 1$ and $0 \le \alpha^2_2 \le 1$
we arrive at
\begin{equation}
|c| \ge | \xi - 1/4|. \label{lowbc}
\end{equation}
Using Eq. (\ref{xi}) one immediately arrives at the inequality
(\ref{bound}) stated in the Introduction. We see that $c$ can
vanish only for $\xi=1/4$, i.e., $(g_1^2 + g_2^2) = 2 e^2$. In
general, when this condition will not be met, we obtain a lower
bound on $c$ depending on the value of $\xi$.

From Eqs.  (\ref{x2coup22}) and (\ref{x2coup}) the two other
couplings $g_{12}$ and $g_{22}$ can be expressed in this notation
as
\begin{equation}           
g_{12} =  - e \alpha_1 \alpha_2 
\left(\sqrt{\frac{\lambda_2}{\lambda_1}} -
\sqrt{\frac{\lambda_1}{\lambda_2}} \right)
\;\;,\;\; g_{22} =  e \frac{\xi}{\sqrt{\lambda_1 \lambda_2}} \;\;.
\label{x2coup2}
\end{equation}

\begin{figure} 
\begin{center} 
\epsfxsize= 7cm {\epsfbox{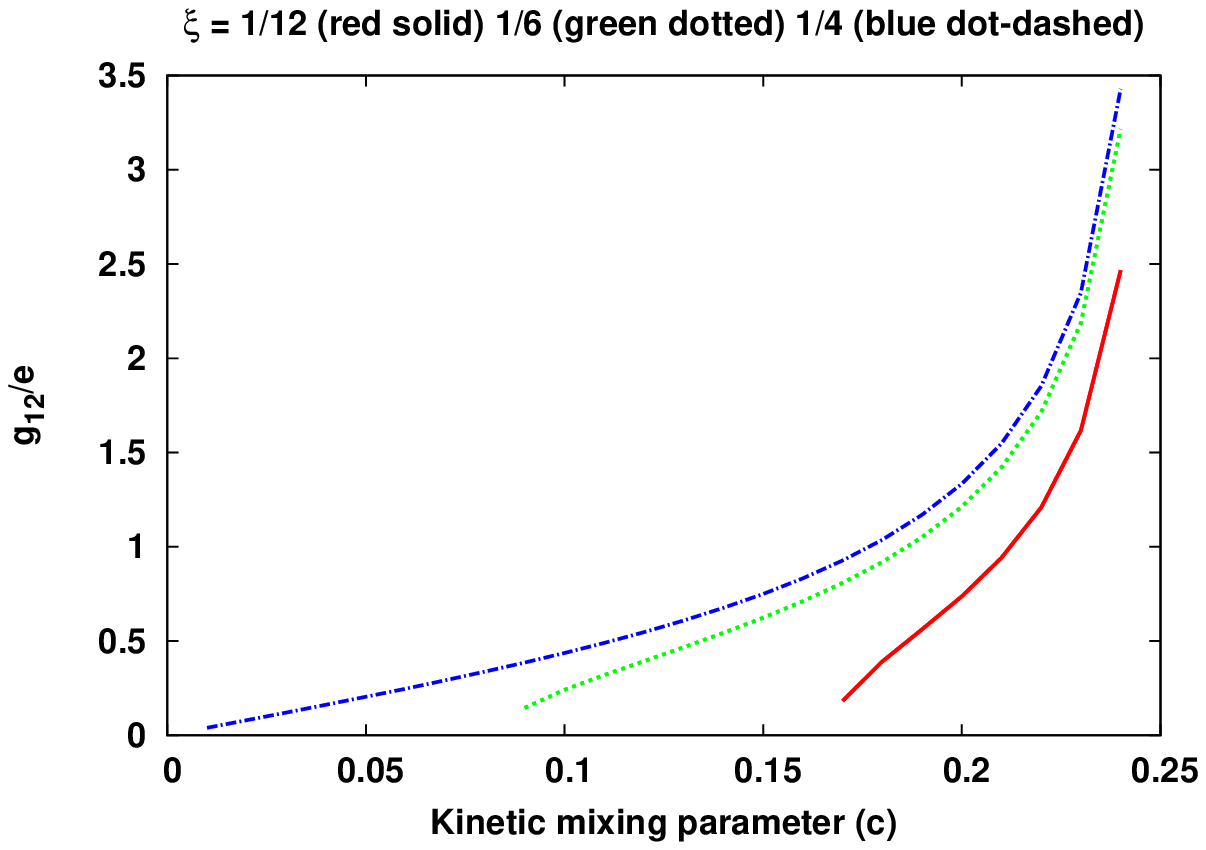}} 
\hskip 1cm
\epsfxsize= 7cm {\epsfbox{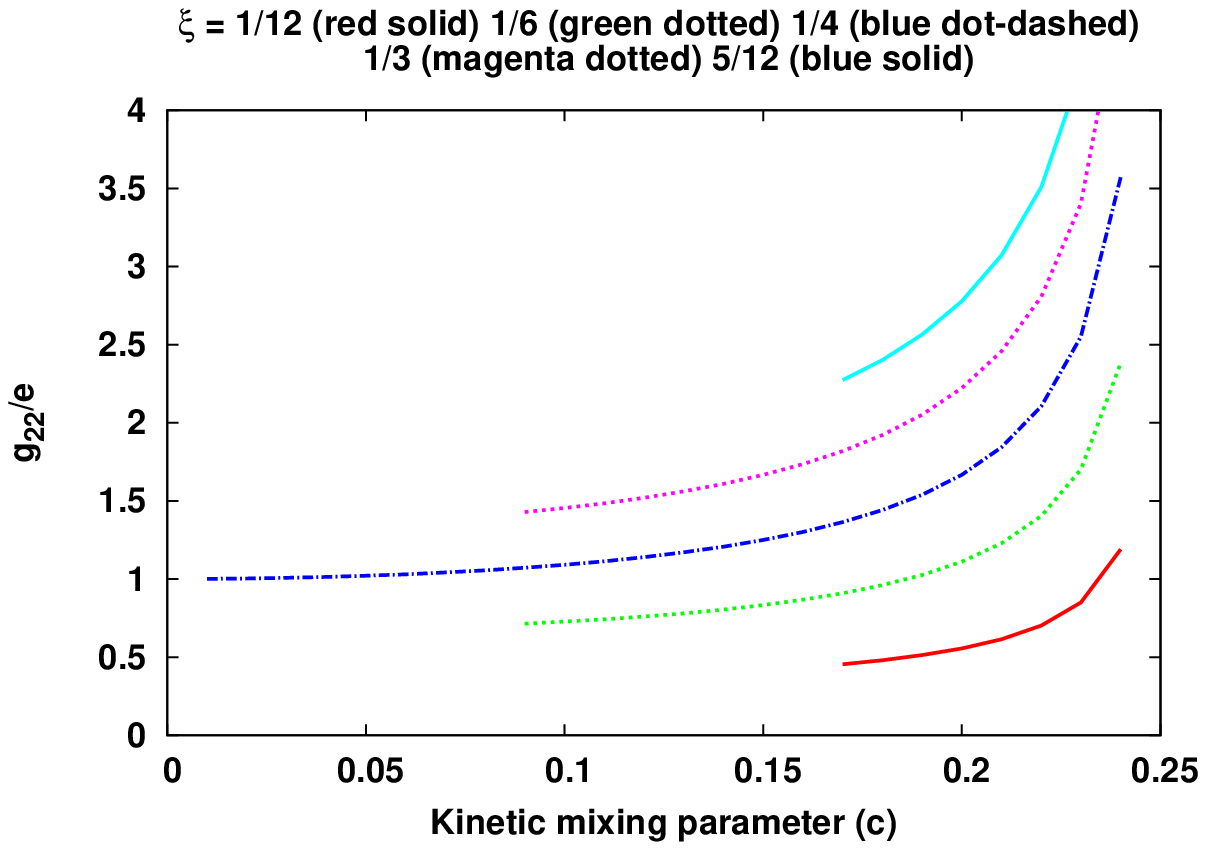}} 
\caption{\small  The couplings $g_{12}$ (left panel) and
$g_{22}$ (right panel) of the dark photon $X_2^\mu$ as a function
of the strength of kinetic mixing $c$ when $g_{11}=e$ and $g_{21}=0$.  In both 
panels curves for
several choices of $\xi$ (indicated in the legend) are shown. The
allowed ranges are $|c| \leq \frac{1}{4}$ and  $0 \leq \xi \leq
\frac{1}{2}$.  However, as $\xi$ increases a more limited range
of $c$ remains consistent.}
\label{coupc} 
\end{center} 
\end{figure}

As is evident from the above discussion both $g_{12}$ and
$g_{22}$ are determined once $c$ and $\xi$ are fixed. Choosing
several values of $\xi$ within the permitted range, we display in
Fig. \ref{coupc} the dependence of $g_{22}$ and $g_{12}$ on $c$.
We have  taken the central value of $\xi=1/4$ and also other
values equidistantly at higher and lower sides of this central
value. We have presented the results for five values of $\xi = $
1/12 (red solid), 1/6 (green dotted), 1/4 (blue dot-dashed), 1/3
(magenta dotted), and 5/12 (blue solid).  It is seen that for
large $c \sim 1/4$ the two couplings are of comparable size.  At
the small $c$ end $g_{12}$ tends to zero while $g_{22}$ tends to
a non-zero limiting value.

Both couplings diverge as $c$ tends towards $1/4$. This is a
reflection of the factor $\sqrt{\lambda_1  \lambda_2}$ in the
denominator in the expressions for $g_{12}$ and $g_{22}$ in Eq.
(\ref{x2coup2}) since from Eq. (\ref{lam}):
\begin{equation}
\lambda_1 \lambda_2 = \frac{1}{16} - c^2 \;\;.
\end{equation} 
Nonethesless physical processes remain finite in the $c
\rightarrow 1/4$ limit  as the mass of $X_2^\mu$ also diverges. Using Eqs.
(\ref{mass2}) and (\ref{alphacr}) one has:
\begin{equation}
m_2^2 = e^2 \xi^2 (Q_1^{s2} + Q_2^{s2})\frac{1}{\lambda_1 \lambda_2} v^2.
\label{mass3}
\end{equation}

For any $\delta$ between 0 and $1/4$ when $\xi$ changes
from $1/4-\delta$ to $1/4+\delta$ $\alpha_1$ and $\alpha_2$ are
exchanged, as can be seen from Eq.   (\ref{alphacr2}).
Because $g_{12}$ depends on the product $\alpha_1 \alpha_2$, the
curves for $g_{12}$ for these cases overlap.  For a given value
of $c$, larger $g_{22}$ corresponds to a higher $\xi$.   Since
$\alpha_{1,2} \leq 1$, we see from Eq. (\ref{alphacr2}) that for
larger values of $\xi$ the kinetic mixing strength $c$ can take
values in a restricted range.  Here we have considered only
positive values of $c$ since, as noted, for negative $c$ one
has $\alpha_1$ interchanged with $\alpha_2$ whereas
$\lambda_1$ and $\lambda_2$ are exchanged. This will take
$g_{12}$ to -$g_{12}$, while $g_{22}$ will be unaffected. 

We observe that once
$\alpha_1$ is fixed by $c$ and $\xi$, the electric charge of a
fermion, $Q$, is given by Eq.  (\ref{chargeQ}) in terms of the
$U(1) \times U(1)$ charges $Q_{1,2}$. The orthogonal charge
combination, $Q'$, is  similarly defined in Eq.   (\ref{chargeQp}).

In the next two subsections we present two  illustrative models.
In the first one $U(1)_{sm} \times U(1)_{dm}$ breaks to $U(1)_{EM}$.
Here suffixes $sm$ and $dm$ indicate visible and dark sectors
respectively whereas the suffix $EM$ denotes electromagnetism. 
In the second example $U(1)_{EM} \times U(1)_{dm}$ breaks to
$U(1)_{EM}$. In the first example gauge bosons of $U(1)_{sm}$
and $U(1)_{dm}$ mix during the spontaneous symmetry breaking
process, whereas in the second case the mixing between
photon and the dark gauge boson is solely due to the
kinetic mixing.

\subsection{Example 1: A toy model for dark matter}
 
In this model there are two sectors, a visible sector denoted
by $U(1)_{sm}$ and a dark sector denoted by $U(1)_{dm}$. 
Even though this model is not realistic as it stands, key features of
our analysis can be demonstrated by this simplified version.
Symmetry breaking is along the following line,
\begin{equation}
U(1)_{sm} \times U(1)_{dm} \longrightarrow U(1)_{EM}
\end{equation} 
To apply this formulation of kinetic mixing to models of dark
matter we consider two classes of particles specified in terms of
the nature of their charges $Q$ and $Q^\prime$.   Of these,
$Q$ corresponds to the current $\hat{J}^\mu_1$ which is
associated with $U(1)_{EM}$. It is a conserved charge
unlike $Q^\prime$ which corresponds to the orthogonal broken
direction. The photon (${X}^\mu_1$) couples through only $Q$
while the dark photon (${X}^\mu_2$) couples to both $Q$ --
coupling $g_{12}$ -- as well as $Q'$ -- coupling $g_{22}$.

\begin{figure} 
\begin{center} 
\epsfxsize= 7cm {\epsfbox{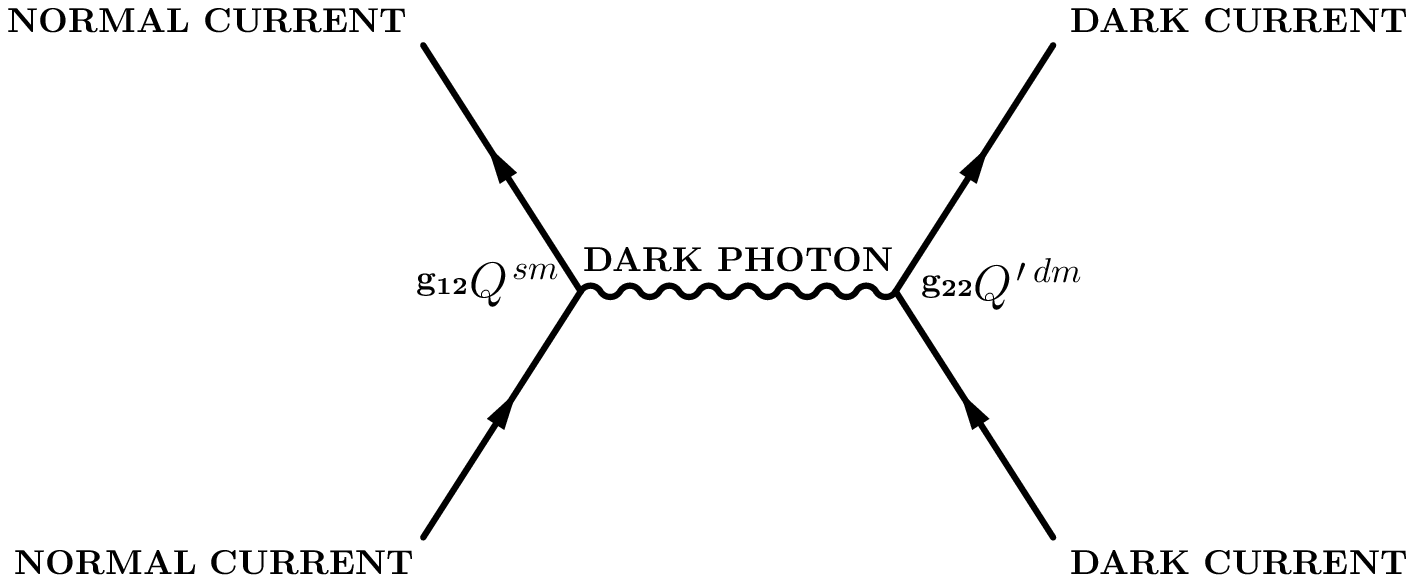}} 
\hskip 1cm
\epsfxsize= 7cm {\epsfbox{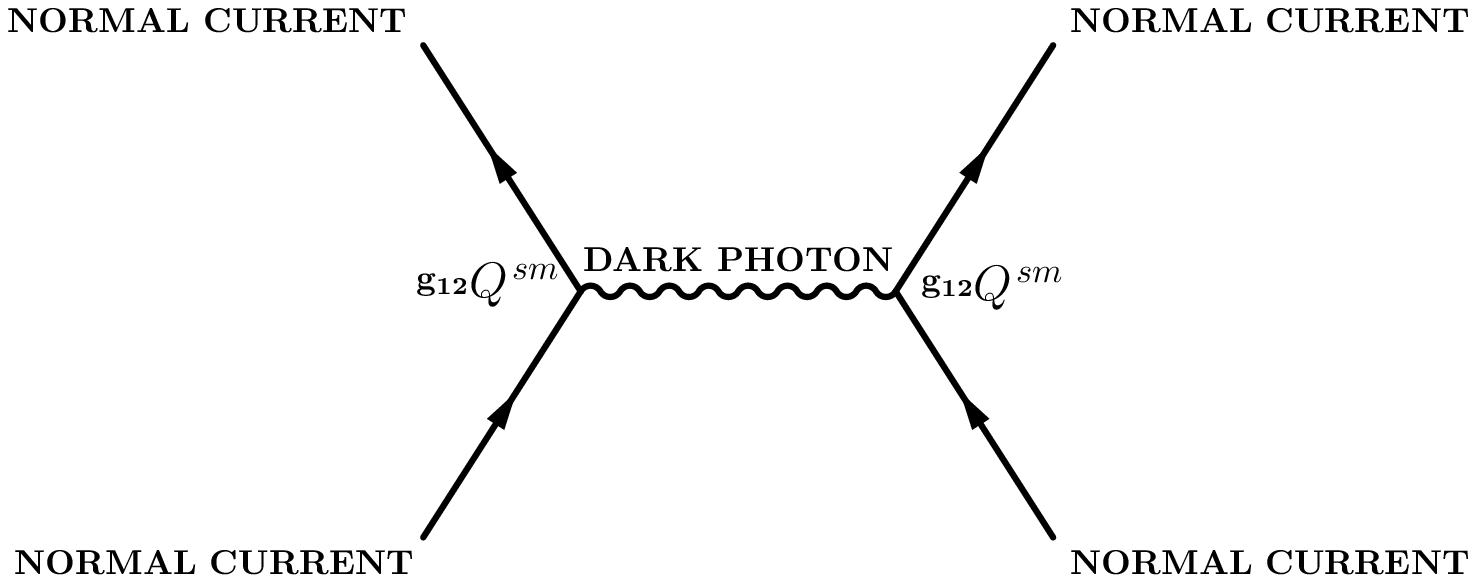}} 
\caption{\small The interactions due to $X_2^\mu$ exchange:
between SM particles and dark matter (left) and between SM
particles themselves (right).}
\label{fint} 
\end{center} 
\end{figure}

There are two classes of particles, namely,  (a) Dark Matter
which is decoupled from the photon by having $Q^{dm} =0$,  and (b) Normal
matter which has $Q'^{sm} = 0$.
By choice, we have the photon coupling only to the SM sector. It
will be of our interest to discuss the coupling of SM with Dark
Matter through the dark photon, $X_2^\mu$, mediated interactions. 
This is shown in the
left panel of Fig. \ref{fint}.  For momentum transfers small
compared to $m_{X^2}$  from
Eqs. (\ref{x2coup2})  the probability amplitude will be
\begin{equation}
{\cal M}_{sm-dm} \propto \frac{g_{12}~Q^{sm} g_{22}~Q'^{dm}}{m_{X^2}^2} = 
- \left[e^2 ~Q^{sm}  Q'^{dm}~ \alpha_1 \alpha_2 
\left(\sqrt{\frac{\lambda_2}{\lambda_1}} -
\sqrt{\frac{\lambda_1}{\lambda_2}} \right)
\frac{\xi}{\sqrt{\lambda_1 \lambda_2}}
\right] \frac{1}{m_{X^2}^2}  \;\;,
\label{amp}
\end{equation}
where $Q^{sm}$ and   $Q'^{dm}$ are respectively the electric
charge of the SM particle and the dark charge of the DM particle.

One can readily extract the dependence of the above
amplitude on $c$. One finds
\begin{equation}
{\cal M}_{sm-dm} \propto \sqrt{16 c^2 - (4\xi - 1)^2} \;.
\label{ampc}
\end{equation}
By a suitable choice
of $\xi$ near the limiting values
\begin{equation}
\xi\rightarrow 1/4 \pm c,
\end{equation}
the right-hand-side of Eq. (\ref{ampc}) can be made
arbitrarily small. Hence, a small and controllable 
interaction cross section
between the standard and dark sectors is a natural consequence of
the model. On the other hand, one may be tempted to
think that for large values of the mixing parameter  $|c| \sim 1/4$
this interaction can be enhanced. However, this will also
modify cross sections of purely standard processes such as
$e^-~e^- \rightarrow e^-~ e^-$ and is very tightly constrained.
For example, the $X_2^\mu$ coupling to SM fermions will result in
interactions within the SM sector as depicted in the right panel
of Fig. \ref{fint}. This leads to the probability amplitude 
\begin{equation}
{\cal M}_{sm-sm} \propto \frac{(g_{12} ~Q^{sm})^2}{m_{X^2}^2} = 
\left[e~Q^{sm}~ \alpha_1 \alpha_2  
\left(\sqrt{\frac{\lambda_2}{\lambda_1}} -
\sqrt{\frac{\lambda_1}{\lambda_2}} \right)
\right]^2 \frac{1}{m_{X^2}^2} \;.
\label{amps}
\end{equation}
The dependence of the above amplitude on $c$ is
\begin{equation}
{\cal M}_{sm-sm} \propto \left(16 c^2 - (4\xi - 1)^2\right) \;.
\label{ampc2}
\end{equation}

Needless to say, one can similarly calculate
scattering within the dark matter sector via $X_2^\mu$-exchange.

\subsection{Example 2: Realistic $U(1)_{dm}$ mixing with QED}\label{example}
 
A simple  and realistic model which appears in the literature of kinetic mixing
is one in which the photon mixes with a $U(1)_{dm}$ gauge
field.  Because the other gauge boson is not yet detected
experimentally, $U(1)_{dm}$ symmetry is broken and the
dark photon is massive. 
Usually the mixing term is  considered as a perturbation and its
effects examined. 

In our approach, which is exact, one must
identify the remaining unbroken symmetry as QED which is also
one of the two initial $U(1)$ symmetries. Thus, one must  demand
$X_1^\mu \equiv A_1^\mu$. From Eqs. (\ref{coup1a}),
and (\ref{chargeQ}) we can write
\begin{equation}
Q=q^1~(\alpha_1-\alpha_2)\cos \phi + q^2~(\alpha_1+\alpha_2)~\sin
\phi \;.
\end{equation}
The requirement that $Q = q^1$ can be achieved by
\begin{equation}
{\rm Either} \;\; \left( \alpha_1 = -
\alpha_2 = {1 \over \sqrt{2}} \;\; {\rm and} \;\; \phi \;
= {\pi \over 4}\right)\;\;{\rm or}\;\; (\phi = 0 \;\; {\rm and} \;\;
\alpha_2 = 0)\;\;.
\end{equation}
Of these, the second option is untenable  as it implies $g_2
= 0$ as a consequence of Eq. (\ref{coup1}).

This result identifies electric charge as the coupling of
one of the factor groups that existed before symmetry breaking.
From Eq.  (\ref{coup1}) it implies 
\begin{equation}
g_1=g_2 \equiv g,~~
~~\tilde g_1 = \frac{g}{2 \sqrt{\lambda_1}},
~~\tilde g_2 = \frac{g}{2 \sqrt{\lambda_2}}\;.
\end{equation}
Thus, in the $A$ basis where gauge bosons have diagonal
couplings with fermions, gauge coupling must be identical
for the two $U(1)$ factors. To the best of our knowledge this
is a new result. Then from Eqs. (\ref{def2}), (\ref{x2coup22}) and
(\ref{x2coup}),
\begin{eqnarray}           
g_{11} &=& g \equiv  e  \;\;,  \\ 
g_{22} &=&  e~\frac{1}{\sqrt{1 - 16 c^2}} \;\;, \\ 
g_{12} &=&  - e~\frac{4 ~c}{\sqrt{1 - 16 c^2}} \;\;.
\label{gij2}
\end{eqnarray}
Using Eq. (\ref{xi}) we get $\xi=1/4$ for which  as shown earlier
$|c| \ge 0$.

Two noteworthy features here are that $g_{22}$, the coupling within the Dark
Matter sector mediated by $X^2_\mu$, is stronger
than normal electromagnetism.
Also Dark Matter couples to ordinary matter via the 
coupling $g_{12}$  which goes to zero as the  kinetic
mixing parameter $c \rightarrow 0$.

Before moving on we would like to draw attention 
to another mode of handling
kinetic mixing that is often used. It is common in the literature
to define the mixing in the basis in which the gauge bosons are
already the mass eigenstates, one of which is massless while the
other has a non-zero mass typically through a St\"{u}ckelberg
mechanism.  In such scenarios the removal of
the kinetic mixing is enabled through the transformation 
\begin{equation}
 \pmatrix{A^1_\mu \cr ~A^2_\mu} = 
 \pmatrix{1 & {-4c \over \sqrt{1 - 16c^2}}     \cr   
0 & {1 \over \sqrt{1 - 16c^2}} }\pmatrix{{X}^1_\mu \cr {X}^2_\mu}.
\label{rsalt}
\end{equation}
Note that this leads precisely to the couplings in eqs.
(\ref{gij2}) for ${X}^1_\mu$ and ${X}^2_\mu$.

\section{Summary and conclusion} 

When a theory has two (or more) $U(1)$ symmetries then the
possibility of gauge kinetic mixing opens up. We have examined
kinetic mixing in a generic model with two $U(1)$ factors where the
symmetry is spontaneously broken as $U^1(1) \times U^2(1)
\rightarrow U^3(1)$. These models are usually
considered 
in the literature using various approaches that commonly
assume a small mixing parameter, $c$, and study physical effects
by varying it. In this paper in contrast we
have focussed on  $c$ without restricting it to
be small. We show that in certain cases the range of 
$c$ is bounded.

Here, as a first step the kinetic mixing term is
removed by an orthogonal rotation and a scaling. It is
convenient to use the charges, $Q_{1,2}$, of fermions and scalars
in this new orthonormal basis to discuss the spontaneous symmetry breaking.
The symmetry breaking identifies a charge, $Q = \alpha_1 Q_1 +
\alpha_2 Q_2$, corresponding to the unbroken gauge symmetry.
 The interactions are then readily expressed in
terms of  $Q$ and an orthogonal charge $Q^\prime$. While the
massless gauge boson couples only to $Q$ (with coupling $g_{11}$)  the
heavy gauge boson has a coupling to  $Q^\prime$ of
strength $g_{22}$ and also to $Q$ given by $g_{12}$. We
derive analytical formulae for these couplings and show that both
$g_{12}$ and $g_{22}$ are controlled by the mixing parameter $c$.
An important result, which can be seen from Fig. \ref{coupc}
is the following.  To be able to identify the
unbroken $U(1)$ coupling with that of electromagnetism  for a
fixed $\xi = (g_1^2 + g_2^2)/8 e^2$ there is a {\em lower} bound on
the magnitude of $c$ given in Eq. (\ref{lowbc}).  The bound
is quoted in a basis where couplings of fermions to gauge bosons
is diagonal. 

As noted, a nonzero $g_{12}$ is responsible for interactions
between the dark and ordinary sectors. The coupling $g_{22}$ leads to
interactions within the dark sector which have been suggested as
an ingredient for the explanation of the halo structure of
satellite galaxies \cite{dmint}. We note that $g_{22}$ need not
be a small coupling unlike $g_{12}$, which is controlled by kinetic mixing.
Such self-interaction is also needed to resolve conflicts between
observation and simulation at the galactic scale and smaller \cite{sps}.
Self-interaction in the dark sector is also needed to explain
signals obtained in the DAMA experiment \cite{dama}.

We have illustrated this theory by two examples related to Dark
Matter. In both cases we have identified the unbroken $U(1)$ as
the electromagnetic group $U(1)_{EM}$.  In the first example,
ordinary matter has only the $Q$ charge, which is now the
electric charge, whereas dark matter has only the $Q^\prime$
charge. The heavy gauge boson is identified with the dark photon
and it couples to visible as well as dark matter. We have shown
the manner in which the coupling of the dark photon to the
ordinary matter depends on the mixing parameter $c$. In the limit
of no kinetic mixing the dark photon does not interact with the
ordinary matter at all (except by gravity) and therefore cannot
be searched easily in scattering experiments. In the second
example we have examined the case where $U(1)_{EM}$ is
kinetically mixed with another $U(1)$. This situation can occur
only when the two gauge groups have same gauge couplings
initially. In this model also we have given analytical formulae
for the coupling strengths of heavy and massless gauge bosons. In
both cases we have derived analytical expressions for the
Dark Matter self-coupling strengths.

The dark photon has been considered here as an intermediary in
interactions linking  dark matter with ordinary matter. There is
also the possibility that a dark photon may be produced on-shell
in physical processes, e.g., in dark matter annihilation.  In the
literature it has been proposed to look for comparatively light
dark photon signals using $e^+e^-$ colliders or electron beam
dump experiments where an emitted dark photon could decay to a
pair of lighter dark matter \cite{detect}. If the dark photon
coupling to dark matter is enhanced to large values by an
appropriate choice of the mixing parameter $c$, as indicated in
sec. \ref{example},  decays to dark matter will become more
prominent.  This will permit the dark photon to be detected
through these proposed tests.

If the dark photon is relatively light, having mass 
around  10
MeV, then it can decay to $e^+e^-$ pairs only, i.e.,
with branching ratio unity,  with a lifetime which goes as
$1/c^2$. Detection of electron-positron pairs with invariant
mass matching the  dark photon mass would be a clear signal.  If
the mass is such that $\mu^+\mu^-$ decays are kinematically
possible then that too could be an alternate detection channel.
As formulated, the dark photon coupling to all SM particles
should be proportional to the respective electric charges. So,
the branching ratio to muons and electrons will differ simply due
to the phase space considerations. Electrons and muons of such
energy can be observed in neutrino detectors, e.g.,
SuperKamiokande. If the dark photons are produced in the
annihilation of much heavier dark matter particles then one can
expect them to be relativistic. In such an event, the decay
products will be collimated in the forward direction. A magnetic
field will help in separating the decay products and also
determine their energy-momentum. A sufficiently high-energy
charged particle, e.g., at an accelerator, will emit dark photons
by bremsstahlung  which, needless to say, will be suppressed
compared to similar $\gamma$ emission by a factor of (${\cal
O}(c^2)$). There are therefore several avenues for testing the
scenario of kinetic mixing discussed in this paper.

{\bf Acknowledgments:} The authors are grateful to Professor E.
Ma for stoking their curiosity on the role of kinetic mixing in
the dark photon models. BB would like to thank E.J. Chun for
discussions. The research of AR is supported by a J.C.
Bose Fellowship of the Department of Science and Technology of
the Government of India. \\


\end{document}